%% file: performance.tex
\documentclass[sigconf,natbib=false]{acmart}
\pdfoutput=1
\usepackage{tikz}

\usepackage[textsize=footnotesize]{todonotes}
\usepackage{listings}
\usepackage{graphicx}
\usepackage{subcaption}
\usepackage{textcomp}
\usepackage{xspace}

\usepackage{url}
\usepackage{multirow}
\usepackage{tabularx}
\newcolumntype{Y}{>{\centering\arraybackslash}X}
\usepackage[binary-units,per-mode=symbol,detect-all]{siunitx}
\DeclareSIUnit\cycle{cy}
\DeclareSIUnit\flop{flop}

\usepackage{xcolor}

\definecolor{codegreen}{rgb}{0,0.6,0}
\definecolor{codegray}{rgb}{0.5,0.5,0.5}
\definecolor{codepurple}{rgb}{0.58,0,0.82}
\definecolor{backcolour}{rgb}{0.95,0.95,0.92}

\lstdefinestyle{mystyle}{
    language=C,
    backgroundcolor=\color{backcolour},
    commentstyle=\color{codegreen},
    keywordstyle=\color{magenta},
    numberstyle=\tiny\color{codegray},
    stringstyle=\color{codepurple},
    basicstyle=\ttfamily\footnotesize,
    breakatwhitespace=false,
    breaklines=true,
    captionpos=b,
    keepspaces=true,
    numbers=left,
    numbersep=5pt,
    showspaces=false,
    showstringspaces=false,
    showtabs=false,
    tabsize=2
}

\title{Model-Based Performance Analysis of the \hyteg Finite Element Framework}
\author{Dominik Th\"onnes}
\orcid{0000-0002-8340-4849}
\email{dominik.thoennes@fau.de}
\affiliation{%
  \institution{Friedrich-Alexander-Universität Erlangen-Nürnberg (FAU)}
  \department{Chair for System Simulation}
  \city{Erlangen}
  \country{Germany}
}

\author{Ulrich R\"ude}
\email{ulrich.ruede@fau.de}
\orcid{0000-0001-8796-8599}
\affiliation{%
  \institution{Friedrich-Alexander-Universität Erlangen-Nürnberg (FAU)}
  \department{Chair for System Simulation}
  \city{Erlangen}
  \country{Germany}
}
\affiliation{%
  \institution{CERFACS}
  \city{Toulouse}
  \country{France}
}

\copyrightyear{2023}
\acmYear{2023}
\setcopyright{rightsretained}
\acmConference[PASC '23]{Platform for Advanced Scientific Computing Conference}{June 26--28, 2023}{Davos, Switzerland}
\acmBooktitle{Platform for Advanced Scientific Computing Conference (PASC '23), June 26--28, 2023, Davos, Switzerland}
\acmDOI{10.1145/3592979.3593422}
\acmISBN{979-8-4007-0190-0/23/06}

\newcommand{\hyteg}{\textsc{HyTeG}\xspace}
\newcommand{\petsc}{\textsc{PETSc}\xspace}
\newcommand{\sympy}{\textsc{SymPy}\xspace}
\newcommand{\pystencils}{\textsc{pystencils}\xspace}
\newcommand{\likwid}{\textsc{LIKWID}\xspace}
\newcommand{\CC}{C\nolinebreak\hspace{-.05em}\raisebox{.4ex}{\tiny\bf +}\nolinebreak\hspace{-.10em}\raisebox{.4ex}{\tiny\bf +}\xspace}
\newcommand{\CPP}{\CC}

\usepackage[acronym]{glossaries}
\newacronym{dof}{DoF}{degrees of freedom}
\newacronym{cl}{CL}{cache line}
\newacronym{LC}{LC}{layer condition}

\RequirePackage[
  datamodel=acmdatamodel,
  style=acmnumeric,
  ]{biblatex}
\begin{document}




\input{abstract}
\maketitle
\input{intro}
\input{application}
\input{petsc}
\input{codegen}
\input{2D-ECM}
\input{multiCore}
\input{conclusion}
\input{acknowledgments}

%\bibliographystyle{ACM-Reference-Format}
%\bibliography{hyteg-perf}
\printbibliography

\end{document}

%% file: abstract.tex
\begin{abstract}
In this work, we present how code generation techniques significantly improve the performance of the computational kernels in the \hyteg software framework.
This HPC framework combines the performance and memory advantages of matrix-free multigrid solvers with the flexibility of unstructured meshes.
The \pystencils code generation toolbox is used to replace the original abstract C++ kernels with highly optimized loop nests.
The performance of one of those kernels (the matrix-vector multiplication) is thoroughly analyzed using the Execution-Cache-Memory (ECM) performance model.
We validate these predictions by measurements on the SuperMUC-NG supercomputer.
The experiments show that the performance mostly matches the predictions.
In cases where the prediction does not match, we discuss the discrepancies.
Additionally, we conduct a node-level scaling study which shows the expected behavior for a memory-bound compute kernel.
\end{abstract}

%% file: intro.tex
%!TEX root = performance.tex

\section{Introduction and Methodology}\label{sec:intro}
This article will deal with the question:
\emph{How do we determine the performance of a program?}
This question has no straightforward answer, but the answer is rather manifold.
In high-performance computing (HPC), the relevant metrics are single-core performance and multi-core performance or scaling.
These aspects must be considered in combination because looking only at scaling tells little about the actual performance of the code and can even be misleading since a slow code will typically scale quite nicely.
The reason is that if shared resources are only utilized to a minor extent, there are more reserves for parallel execution.
If the shared resources are already highly utilized by a sequential program, achieving good scaling is harder.
This paper lays out the basis for performance engineering by thoroughly analyzing the computational kernels of the \hyteg \anon{\cite{kohl2019hyteg}} simulation framework, specifically looking at the single-core performance.

\subsection{Sparse Matrices}
\label{sec:matrix}

Operations on matrices are a fundamental building block in numerical algorithms, and they often form the basis for solving computational problems where matrices with huge dimensions can be involved.
Generally, a matrix is a rectangular array arranged in rows and columns.
These \emph{dense} matrices quickly reach the memory limits.
For example, if a problem consists of $10^8$ unknowns, a dense matrix that describes the connection between these points would reach a size of $10^8\cdot10^8*\SI{8}{\byte} = \SI{80}{\peta\byte}$, which is more than any supercomputer can provide.
\emph{Sparse matrices} form an important subclass when most elements are zero.
This sparsity is exploited by omitting the zero entries when storing the matrix, which can dramatically reduce the amount of memory required.
The simplest version uses a list where each element contains the row, column, and value of a matrix entry (Coordinate list (COO)).
The memory footprint can be further optimized using a compressed format like the \emph{compressed row storage (CRS)}\cite{barrett1994templates} format.
The CRS uses a list to store the column indices of each matrix entry in the matrix.
The row indices of the matrix entries are not stored explicitly but only at the start of a new row in the column list.
The memory required for the column is reduced from $\#matrix\_entries$ to $\#matrix\_columns$ using the compressed format.

\subsection{\textbf{Hy}brid \textbf{Te}trahedral \textbf{G}rids}
\label{sec:hyteg}

\begin{figure}
  \centering
  \begin{subfigure}[t]{0.49\columnwidth}
      \includegraphics[width=\textwidth]{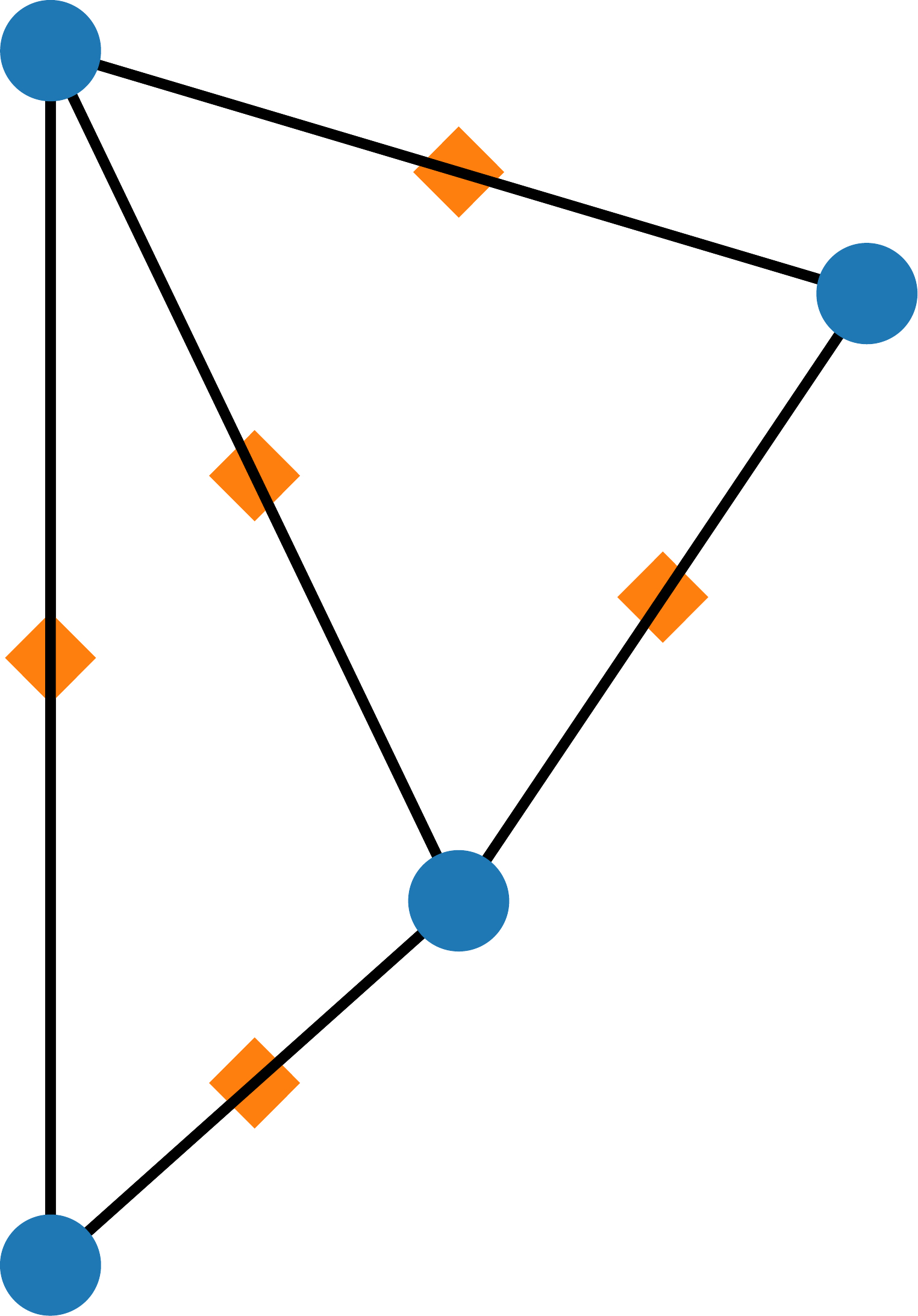}
      \caption{Initial mesh}
      \label{fig:refinementCoarse}
  \end{subfigure}
  \hfill
  \begin{subfigure}[t]{0.49\columnwidth}
      \includegraphics[width=\textwidth]{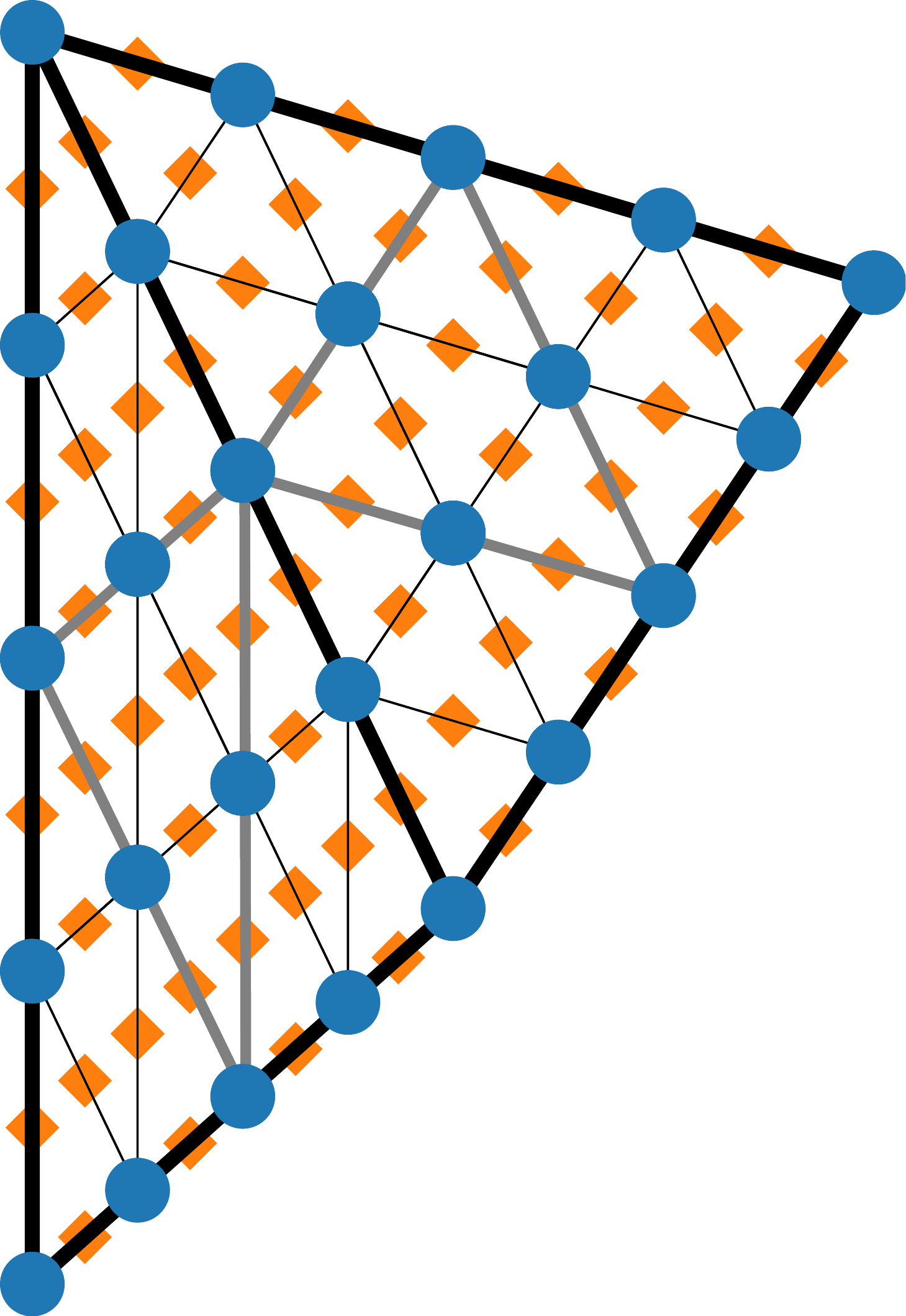}
      \caption{Mesh after two refinement steps. The first refinement level is shown in grey.}
      \label{fig:refinementFine}
  \end{subfigure}
  \caption{Refinement process for a mesh consisting of two triangles. Degrees of freedom are shown as blue circles on vertices and orange rectangles on edges.}
  \label{fig:refinement}
\end{figure}

The \hyteg software framework was developed in the \anon{TERRANEO project \cite{bauer2020terraneo}} as
successor of HHG \cite{bergen2004hhg}.
The fundamental concept is to use an unstructured grid of triangles or tetrahedra and perform a uniform grid refinement in each of these.
This concept combines the advantages of unstructured grids for mesh flexibility and structured grids for performance.
The refinement process leads to a structured grid within each triangle.
Figure~\ref{fig:refinement} illustrates the process with Figure~\ref{fig:refinementCoarse} showing the initial mesh and Figure~\ref{fig:refinementFine} the mesh after two refinement steps.

This structure enables the usage of matrix-free methods and provides a natural basis for multigrid methods.
As in multigrid terminology, we refer to the meshes obtained from the refinement steps \emph{levels}.
For example, a grid on level five is constructed by performing five refinement steps.
In this article, we focus on two-dimensional grids only, even though \hyteg is capable of three-dimensional computations as well.

The framework utilizes \gls{dof} located on the vertices and the edges of the triangles.
Figure~\ref{fig:refinement} shows blue circles for \gls{dof} located at the vertices and orange diamonds for \gls{dof} at the edges.
This arrangement allows, for example, to use finite elements with a $\mathbb{P}_1$ or $\mathbb{P}_2$ discretization.

\begin{figure}
  \centering
  \includegraphics[width=0.8\columnwidth]{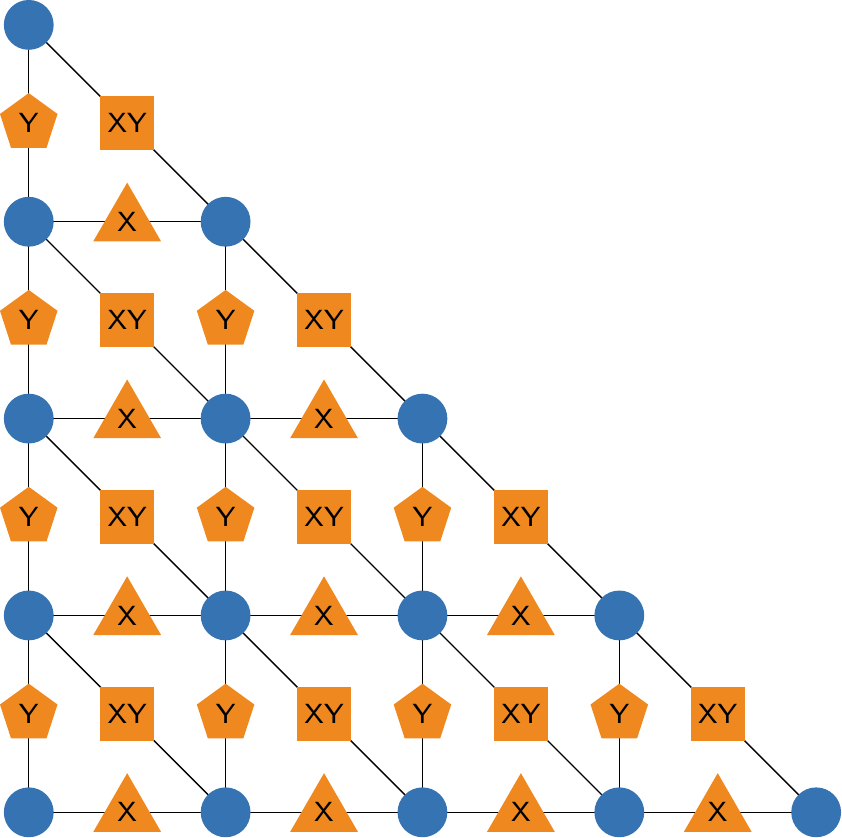}
  \caption{Grouping of degrees of freedom. Each type has exactly the same neighborhood except for the boundaries}
  \label{fig:dof-groups}
\end{figure}

One technical detail worth mentioning here is that the connection to all neighbors is alike in the case of \gls{dof} located at the vertices.
This is not the case for \gls{dof} located at the edges.
In contrast, there are three groups for each side of the triangle.
We refer to these groups as X, Y, and XY, as shown in Figure~\ref{fig:dof-groups}.
The connections to the neighboring \gls{dof} are again the same within one of these groups.
These connections are also known as \emph{stencils}, which contain the corresponding weights to couple one \gls{dof} to its neighbors.
In an \emph{unstructured grid}, these stencils are typically different from each other.
A structured grid is created within each triangle by using regular refinement.

The similarity within each group of \gls{dof} means that the stencils, i.e.~the matrix rows,  are also identical for each group.
Therefore, only four individual stencils are needed within each triangle, one for the vertex \gls{dof} and one for each of the different groups of edge \gls{dof}.
In contrast to a matrix that explicitly describes and stores the connections between all \gls{dof},
a so-called \emph{matrix-free} storage is naturally obtained when each stencil is constant.

\subsection{Execution-Cache-Memory Model}
\label{sec:ecm-model}

This part introduces the performance model used in Section~\ref{sec:performance-model}.
For additional information, see, e.g., \cite{stengel2015layercondition}. % as a starting point.
The ECM model can be % described
viewed as an extension of the Roofline model \cite{williams2009roofline}, which similarly uses the principle that the execution speed of a program can be limited by either data transfer or the execution of operations.
Both models also assume that these two limiting factors perfectly overlap, meaning data transfer does not influence the execution and vice versa.

In contrast to the Roofline Model, the ECM model uses processor cycles as a time unit, which has the advantage that the predictions are mostly independent of the processor frequency.
Additionally, the model can take overlapping and non-overlapping parts into account, which is essential to predict
in-core execution times.

In-core describes the actual computation in the processor and the loads and stores to and from the registers but leaves out the transfers within the caches.
The in-core contribution is split into two parts:
The first one is called $T_{OL}$, which is short for $T_{OverLapping}$
since the computation itself does fully overlap with data transfers.
The loads and stores from and into the registers,
however, do not overlap with cache transfers and are therefore named $T_{nOL}$ for $T_{non OverLapping}$.
To determine $T_{nOL}$ and $T_{OL}$, the Intel Architecture Code Analyzer (IACA) \cite{iaca} can be used.
This tool analyses the kernel's C++ source code or assembler code and predicts the execution happening within the CPU.

The transfers between the cache levels are also accounted for as $T_{data}$.
Splitting $T_{data}$ into different parts for each interface between two adjacent cache levels or the main memory which leads to, e.g., $T_{data} = T_{L1L2} + T_{L2L3} + T{L3MEM}$ for a CPU with three levels of caches.
The unit for data transfer used by the ECM model is typically one \gls{cl} since this is the smallest data unit that can be transferred in the caches.
One \gls{cl} on modern Intel CPUs has a size of \SI{64}{\byte}.
A \emph{work unit} is used for the computation, which includes all the instructions necessary to process all elements in one \gls{cl}.

For example, eight iterations would be used as a work unit if an analyzed loop adds two vectors using double-precision elements.
This assumes the typical double-precision element size of \SI{8}{\byte}, which means eight elements are contained in one \gls{cl}.

The ECM is often displayed in a short form:\\
 $T_{OL} || T_{nOL} | T_{L1L2} | T_{L2L3} | T_{L3MEM}$\\
The prediction for the case that the dataset fits into a certain cache level or the main memory is determined by adding up the respective transfer time. For example, if the data is in the L3 cache: $T_{nOL} + T_{L1L2} + T_{L2L3}$.
The actual prediction is the maximum of $T_{OL}$ and the sum of the data transfers.

A crucial tool when modeling performance is the so-called \textit{\gls{LC}} \cite{stengel2015layercondition}, which predicts the caching behavior of stencil codes.
These kernels typically apply the stencil at each grid point inside loops that span all dimensions.
Each iteration updates points by accessing various neighbors.
In a two-dimensional grid, the points are typically sequential in memory for one dimension, but there are offsets in the other.
An earlier iteration might have accessed some of the neighboring points already, which means these points might still be present in the cache.
At least one point of the stencil has never been accessed before and is loaded for the first time.
The vertical distance of the stencil also called the height, defines how many rows of the grid array are required to apply the stencil.
If we assume sequential memory in the horizontal direction, a stencil that accesses the top and bottom neighbor has a height of three rows.
The layer condition states that if the height of the stencils times the memory for one row is smaller than the cache size, then only the elements that have never been accessed are cache misses.

\subsection{Hardware Description}\label{sec:machine}

% Throughout this work, an Intel Skylake Xeon Platinum 8174 CPU is used, which is installed in the SuperMUC-NG\footnote{https://www.lrz.de/services/compute/supermuc/supermuc-ng/} supercomputer.
Throughout this article, we will use an Intel Skylake Xeon Platinum 8174 CPU, as it is currently used in the
SuperMUC-NG \cite{supermuc-ng} supercomputer.
This CPU supports both a \SI{205}{\watt} and \SI{240}{\watt} operational TDP mode.
For SuperMUC-NG, the TDP is set to \SI{205}{\watt}, which results in a nominal CPU frequency of \SI{2.7}{\giga\hertz}.

\begin{table}
  \begin{tabular}{@{}cccc@{}}
    \toprule
    load streams & store streams & bandwidth & cycles per \\
    & & &  cache line\\
    \midrule
    1 & 1 & \SI{70}{\giga\byte\per\second} & $2.5$\\
    3 & 1 & \SI{87}{\giga\byte\per\second} & $2.0$\\
    1 & 3 & \SI{60}{\giga\byte\per\second} & $2.9$\\
    \bottomrule
  \end{tabular}
    \caption{Bandwidth of different kernel using the Intel Skylake Xeon Platinum 8174 CPU at \SI{2.7}{\giga\hertz}.}
    \label{tab:bandwidth}
    \vspace{-10mm}
\end{table}

On the Skylake architecture, one can use Advanced Vector Extensions 2 (AVX2) instructions, which means that the equivalent of four \verb-mul- or \verb-add- instructions can be executed in a single instruction when using double precision.
Furthermore, fused multiply-add instructions (FMA) are also available, which would also execute the AVX \verb-mul- and AVX \verb-add- in the same instruction.
The Skylake architecture can perform two FMA AVX (\verb-vfmadd*-)instructions per cycle \cite{inst-table}.
It is worth noting that AVX-512 is also supported, but the clock frequency will be reduced when using AVX-512-enabled code.
This leads to the fact that mainly kernels that are bounded by the computation and not the data transfer will benefit.
Since this is not the case for the kernels analyzed in this paper, AVX-512 does not yield a performance benefit and is, therefore, not enabled.

Several tests were conducted on the main memory bandwidth using the \likwid tool suite.
The results for different combinations of load and store streams are presented in Table~\ref{tab:bandwidth}.
It is important to note that the number of streams is the actual number of streams from an application point of view.
This implies that the store stream also includes one load stream for write-allocate.

In contrast to the roofline model, the ECM also requires the quantities of transfers between the caches.
Table~\ref{tab:supermuc} lists some of the relevant specifications of this processor and the cache-hierarchy characteristics that are used as a basis for the performance modeling.

For the bandwidth of the caches, \emph{half-duplex} and \emph{full-duplex} mean that the cache can either read or write (half-duplex) or that both can happen simultaneously (full-duplex).
In the latter case, if reading and writing co-occur, the actual bandwidth would be $2 \cdot 16$ B/cy.
Normalized to one \gls{cl}, which has \SI{64}{\byte}, it takes one cycle to transfer between L1 and L2 and four cycles between L2 and L3.

\begin{table}
\begin{tabular}{lr}
  \toprule
  cache line size & \SI{64}{\byte} \\
  L1 cache size & \SI{32}{\kibi\byte}   (per core)\\
  L2 cache size & \SI{1024}{\kibi\byte} (per core)\\
  L3 cache size & \SI{33}{\mebi\byte}   (shared)\\
  Main Memory & \SI{96}{\giga\byte}\\
  Clock Speed (fixed) & 2.7 GHz\\
  Cores & 24\\
  Bandwidth L1 $\leftrightarrow$ L2 & \SI{64}{\byte\per\cycle} (half-duplex)\\
  Bandwidth L2 $\leftrightarrow$ L3 & \SI{16}{\byte\per\cycle} (full-duplex)\\
  \bottomrule
\end{tabular}

\caption{Intel Skylake Xeon Platinum 8174 CPU specification and cache characteristics that are used as a basis for the performance modeling. For half-duplex, only one direction at a time can achieve the maximal bandwidth, while full-duplex can use maximal bandwidth in both directions simultaneously.}
  \label{tab:supermuc}
\end{table}

In addition to the transfers between different cache levels, the bandwidth between the L3 cache and the main memory is required.
To calculate the cycles it takes to get one \gls{cl} from L3 cache to memory, the bandwidths from Table\ref{tab:bandwidth} are used and converted.
For one load and store stream for example this reads $ (\SI{64}{\byte} \cdot \SI{2.7}{\giga\hertz}) /\SI{70}{\giga\byte\per\second} = \SI{2.47}{\cycle}$.

\subsection{Outline}
The target application in developing \hyteg is simulating convection in the outer earth mantle.
This can be modeled by a buoyancy-driven convention in Stokes flow.
Section~\ref{sec:app} shows a simple example of this application.

As mentioned before, the possibility to use matrix-free methods is one of the defining features of \hyteg.
A general comparison of matrix-free vs. sparse matrix methods is drawn in Section~\ref{sec:petsc}, as well as some experiments comparing \hyteg with the popular scientific software suite \petsc \cite{balay1997petsc,balay2019petsc} regarding memory consumption.

To combine the benefits of maintainability and performance \hyteg uses code generation for computationally intensive kernels.
The process of how these kernels are created and integrated is described in Section~\ref{sec:codegen}.

In Section~\ref{sec:performance-model}, the Execution-Cache-Memory (ECM) model \cite{treibig2010ecm,hager2016ecm} is used to analyze the performance of the kernels in \hyteg that execute the matrix-vector multiplication.

Finally, an intra-node scaling experiment of the investigated kernel is shown in Section~\ref{sec:multicore}, followed by the conclusion and outlook in Section~\ref{sec:outlook}.

%% file: application.tex
%!TEX root = performance.tex
%
\section{Plume in Rectangular Domain}\label{sec:app}

\begin{figure}
    \centering
    \begin{subfigure}[t]{0.2\textwidth}
        \includegraphics[width=\textwidth, height=200pt]{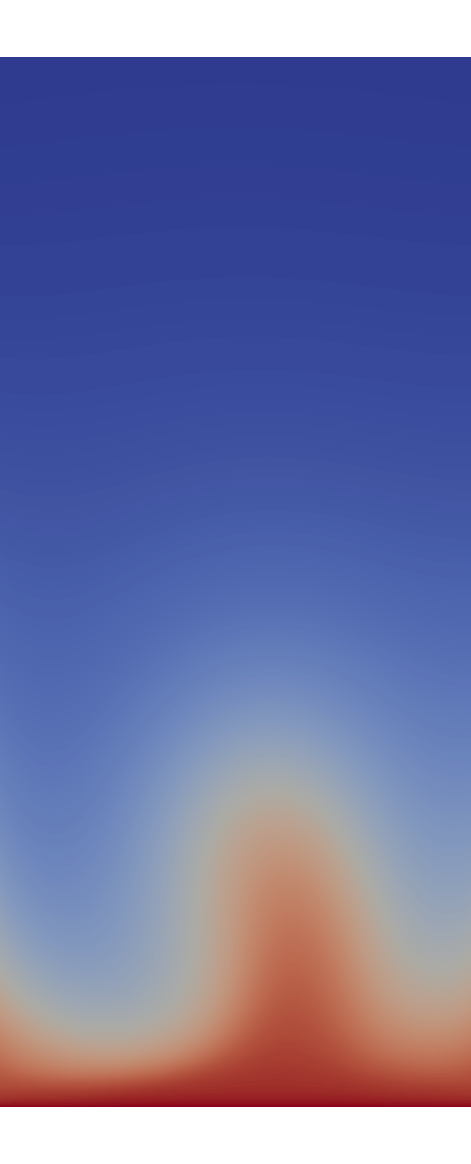}
        \caption{10 time steps}
    \end{subfigure}
    \hfill
    \begin{subfigure}[t]{0.2\textwidth}
        \includegraphics[width=\textwidth, height=200pt]{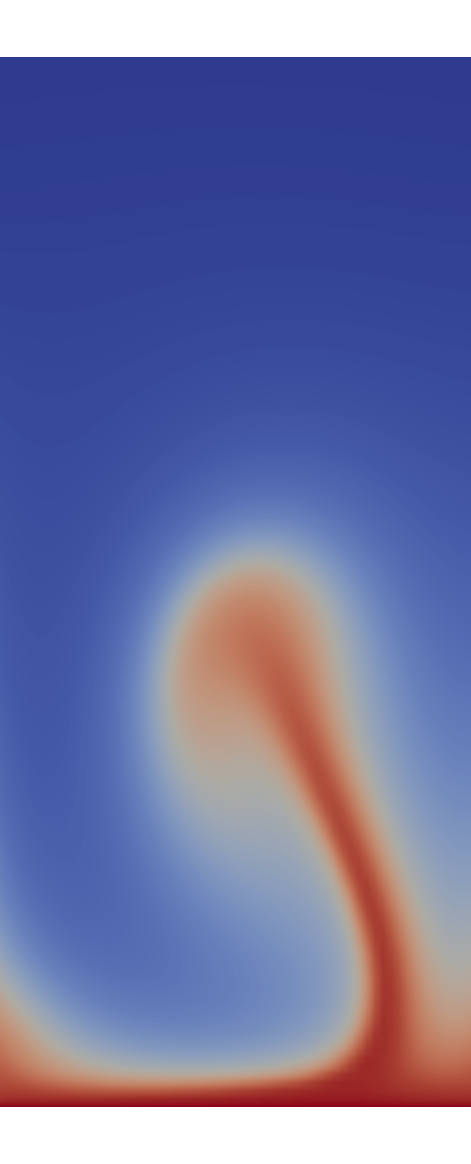}
        \caption{30 time steps}
    \end{subfigure}

    \begin{subfigure}[t]{0.2\textwidth}
        \includegraphics[width=\textwidth, height=200pt]{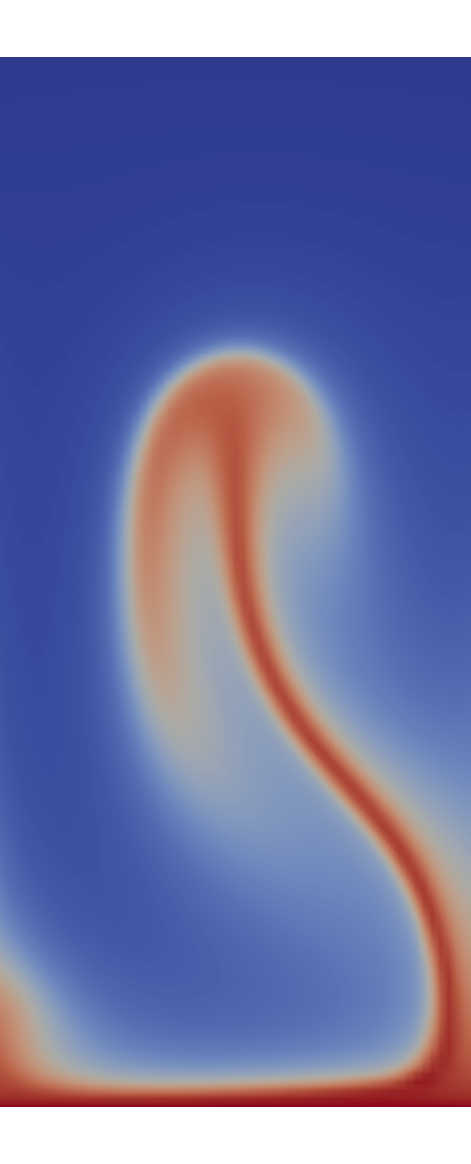}
        \caption{50 time steps}
    \end{subfigure}
    \hfill
    \begin{subfigure}[t]{0.2\textwidth}
        \includegraphics[width=\textwidth, height=200pt]{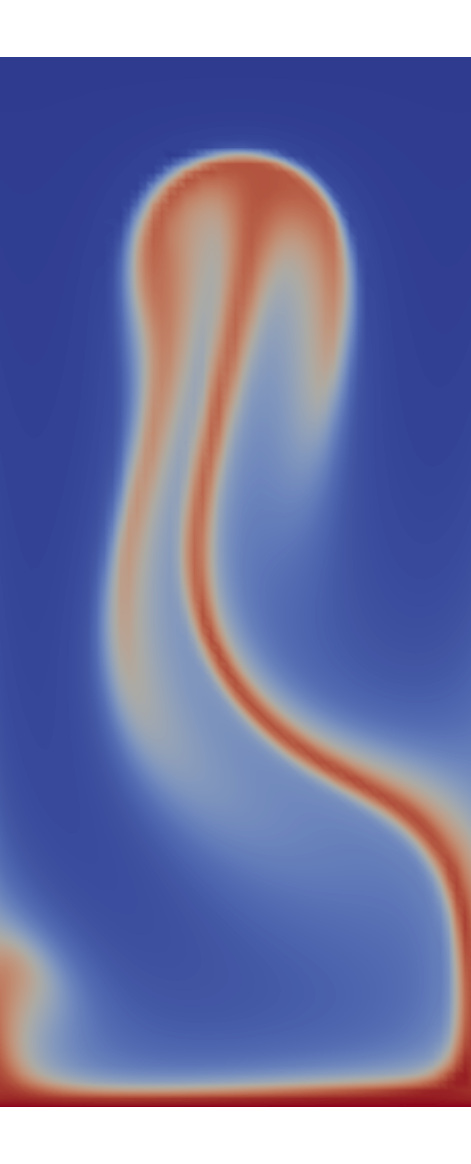}
        \caption{70 time steps}
    \end{subfigure}
    \caption{Buoyancy-driven convection in Stokes flow discretized with Taylor-Hood ($\mathbb{P}_2 \mathbb{P}_1$) elements and a particle-based characteristics method.}
    \label{fig:2dplume}
\end{figure}

To demonstrate the capabilities of the \hyteg software framework, we show the simulation of buoyancy-driven convection in Stokes flow, which is described by an advection-diffusion problem.
We use Taylor-Hood ($\mathbb{P}_2 \mathbb{P}_1$) elements for the discretization of the convection and a particle-based characteristics method for the advection.
\anon{\cite{kohl2022mmoc}} presents details about the particle-based characteristics method and its solvers.
The convection is solved using a MINRES (\cite{paige1975minres}) solver, which is easier to compare to \petsc than a more complex multigrid solver.
Figure~\ref{fig:2dplume} shows a visualization of the simulation for different time steps.
To obtain a baseline for the performance analyses, we profiled the application on refinement level eight with our na\"ive \CPP version of the kernels without using the generated kernels described in Section~\ref{sec:codegen}.
The total number of \gls{dof} is $1.1 \cdot 10^{6}$ in this example application.
When looking at the MINRES solver used to solve the convection, the matrix-vector multiplication accounts for over \SI{57}{\percent} of the runtime, which makes it a good candidate for optimization.

%% file: petsc.tex
%!TEX root = performance.tex
\section{Comparing matrix-free against sparse matrix methods}
\label{sec:petsc}
This section analyzes the performance benefits of matrix-free methods against sparse-matrix methods based on the two-dimensional \hyteg grids described in Section~\ref{sec:hyteg}.
A sparse matrix format stores all entries of the matrix except the ones that are zero.
Additionally, the positions of the entries in the matrix are also stored.
On the other hand, matrix-free methods do not store the entries, drastically reducing memory usage and therefore increasing the performance.
Each row of the matrix describes the connection to the neighbors by storing a factor for each neighboring point.
When updating a point, the corresponding sparse matrix row has to be loaded.
Depending on the matrix's size, the data transfers occur from the caches or the main memory.
Due to the regular refinement in \hyteg as described in Section~\ref{sec:hyteg}, the rows representing the points inside one of the triangles are identical.
In matrix-free methods, the row is often referred to as \verb-stencil- and the factors as \verb-weights-.
If the stencil weights are constant, they can reside in the registers or the cache, resulting in the advantage in memory transfers mentioned above.
However, if the weights vary, using matrix-free methods is more complex.
It is still possible to compute the weights on the fly to reduce the memory transfers, but this introduced computational overhead compared to non-matrix-free methods.
The number of floating-point operations that need to be performed is identical.
Therefore, this aspect is neglected in our analysis.
At first, we analyze the differences theoretically, where the data volume for matrix-free and non-matrix-free variants is determined and compared.
In the second step, a benchmark application compares the actual \hyteg routines against \petsc, one of HPC's most popular libraries for matrix operations.

\subsection{Theoretical analysis}
\label{sec:theory}

\begin{figure}
  \centering
  \begin{subfigure}[t]{0.45\columnwidth}
      \includegraphics[width=\textwidth]{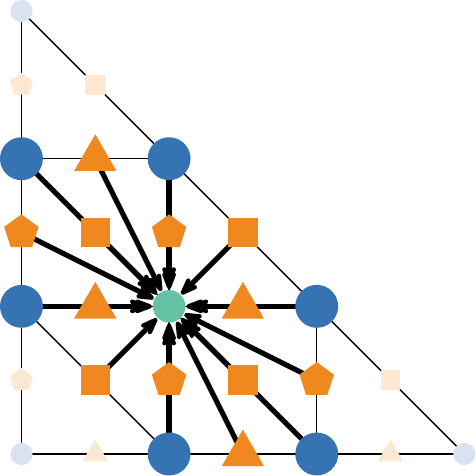}
      \caption{Stencil for one vertex depending on vertices and edges}
      \label{fig:to-vertex-stencil}
  \end{subfigure}
  \hfill
  \begin{subfigure}[t]{0.45\columnwidth}
      \includegraphics[width=\textwidth]{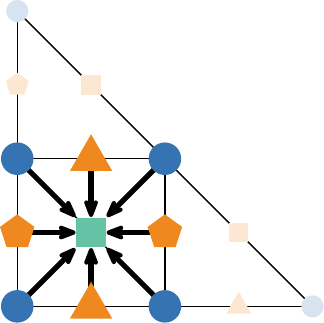}
      \caption{Stencil for one group of edges (XY) depending on vertices and edges}
      \label{fig:to-xyedge-stencil}
  \end{subfigure}
  \caption{Stencils for the vertices and one group of edges. The other edge groups have the same number of neighbors or stencil weights.}
  \label{fig:stencils}
\end{figure}

Since \hyteg performs a regular refinement inside each element, the total number of \gls{dof} can be determined depending on the refinement level.
The number of \gls{dof} that are located on the vertices of the grid is
$$
\#vertex = \frac{(2^l + 1) * (2^l + 1 + 1)}{2}
$$
where $l$ is the refinement level.
Figure \ref{fig:refinement} shows an example mesh for refinement levels zero and two.
This is slightly different for the \gls{dof} located at the edges.
Due to the regular refinement, the edges can be grouped into three different orientations. Each orientation corresponds to one side of the original triangle as shown in Figure~\ref{fig:dof-groups}.
Each of the three orientations can be treated identically and named X, Y, and XY in this paper.
The number for \textbf{one single} orientation ($x$ in this case) is
$$
\#edge = \frac{(2^l) * (2^l + 1)}{2}
$$.
Since there are three types of edge orientations, the total number of \gls{dof}, including edges and vertices, is:
$$
\#\texttt{dof} = \#vertex + 3 \cdot \#edge\texttt{.}
$$
As stated, $\#\texttt{dof}$ is the same for matrix-free and non-matrix-free algorithms.
The difference between the two occurs when considering the stencil weights.
Figure \ref{fig:stencils} shows the stencils for a vertex (\ref{fig:to-vertex-stencil}) and the XY group of edges (\ref{fig:to-xyedge-stencil}).
To get the total number of stencil entries, one has to sum up all stencil weights corresponding to the neighboring vertex and edge \gls{dof}.
For the \gls{dof} located at the vertices, this sums up to 19 entries.
Seven correspond to the neighboring vertices (including the center), and twelve correspond to the adjacent edges of various groups.
In the case of \gls{dof} located at the edges, four neighboring vertices and four adjacent edges must be considered.
The entries sum up to a total number of nine stencil weights, including the edge itself.

Table~\ref{tab:memory-level-10} displays the required memory for a single triangle after ten refinement steps, as one example.
The calculations neglect the domain boundary where the stencils are not fully populated and assume double precision entries with a size of \SI{8}{\byte}.

\begin{table}
  \begin{tabular}{lcccr}
    \toprule
    $\#vertex$ & &&= & 525825 \\
    $\#edge$ & &&= & 524800 \\
    $\#\texttt{dof}$ & &&= & 2100225\\
    \midrule
    $\texttt{dof}\_mem$ & =& $\SI{8}{\byte} \cdot \#\texttt{dof}$ & =  & $\SI{16.8}{\mebi\byte}$\\
    $vertex\_mem$ & = & $\SI{8}{\byte} \cdot 19 \cdot \#vertex$ & = & $\SI{80.0}{\mebi\byte}$ \\
    $edge\_mem$ & = & $\SI{8}{\byte} \cdot 9 \cdot 3 \cdot \#edge$ & = & $\SI{113.4}{\mebi\byte}$ \\
    \bottomrule
  \end{tabular}

  \begin{tabular}{lr}
    \toprule
    memory mat-vec sparse matrix & \SI{332.1}{\mebi\byte} \\
    \midrule
    memory mat-vec theoretical matrix-free &  \SI{33.6}{\mebi\byte}\\
    memory mat-vec \hyteg implementation & \SI{67.2}{\mebi\byte}\\
    \bottomrule
  \end{tabular}

  \caption{Required memory for degrees of freedom and stencils if a uniform refinement is applied ten times (refinement level 10). $vertex\_mem$ refers to the memory required to store all vertex stencils and $edge\_mem$ to store all edge stencils, respectively.}
    \label{tab:memory-level-10}
  \end{table}

%dof_memory = 16801800
%vertex_stencil_memory = 79925400
%edge_stencil_memory = 113356800
When summing up the stencil entries, this means that for storing the matrix \SI{80.0}{\mebi\byte} + \SI{113.4}{\mebi\byte} = \SI{193.4}{\mebi\byte} are needed.
Additionally, the position of each stencil entry in the whole matrix needs to be stored.
A compressed format is often used in scientific computing to reduce the memory footprint of sparse matrices.
When using a compressed row or column format, the additional data is one index for each element and one for each row/column in the matrix.
Using 32-bit integers this leads to $\SI{193.4}{\mebi\byte} \cdot 0.5 = \SI{96.7}{\mebi\byte}$ and $\SI{16.8}{\mebi\byte} \cdot 0.5 = \SI{8.4}{\mebi\byte}$ of additional indexing memory.
In summary, the storage needed for the whole matrix is \SI{298.5}{\mebi\byte}.
The source and destination vectors allocate \SI{16.8}{\mebi\byte} each.
The total memory footprint for the matrix-vector multiplication with non-matrix-free matrix storage is, therefore, \SI{332.1}{\mebi\byte}.
In comparison, a matrix-free algorithm only requires the source and the destination vector, i.e., \SI{33.6}{\mebi\byte}.
Therefore the sparse matrix needs a factor of $\frac{332.1}{33.6} = 9.9$ more memory in this scenario.

\subsection{Comparing implementations}\label{sec:comparing}
In this part, we compare the \petsc library, which assembles the full matrix, against implementing a matrix-free algorithm in the \hyteg framework.
This na\"ive implementation in \hyteg is optimized for usability, using various abstractions provided by modern C++ programming capabilities.
It is not optimized to reach the highest possible performance but primarily to use a matrix-free structure.

One specialty about the implementation in \hyteg is that the matrix-vector multiplication for discretizations that involve \gls{dof} at vertices and edges is split into four different kernels.
Each of these kernels handles a different combination: vertex to vertex, edge to vertex, vertex to edge, and edge to vertex.
Using these different kernels was the most straightforward approach when moving from vertex-only meshes to meshes that support \gls{dof} at the vertices and the edges.
One possible optimization is merging the four kernels into one to reduce the number of data streams.
Because \hyteg splits the matrix-vector multiplication into four parts, the memory has to be loaded multiple times.
In Figure~\ref{fig:stencils}, this separation can be recognized by the difference in color.
Figure~\ref{fig:to-vertex-stencil} and Figure~\ref{fig:to-xyedge-stencil} shows two kernels each.
One kernel considers only the blue vertices, and the other only the orange edges.
All kernels update the green points.

Both the \gls{dof} located at the vertices and the \gls{dof} situated at the edges must be loaded two times and stored two times.
This amounts to $4 \cdot \SI{16.8}{\mebi\byte} = \SI{67.2}{\mebi\byte}$.
Therefore, the expected difference is not a factor of $9.9$ as derived in Section \ref{sec:theory} but rather $\frac{332.1}{67.2} = 4.9$.

To confirm the calculations, we used \likwid to measure the main memory volume during the execution of the application.
The main memory volume denotes the amount of memory loaded from and stored to the main memory during the computation.
Ideally, one would instead measure the peak amount of memory allocated instead of the total volume transferred during the execution.
However, measuring the process's peak amount of memory allocated is a nontrivial task.
There are different methods like the Linux function \verb-getrusage-, internal methods of the \petsc library, or external tools like Valgrind\footnote{https://valgrind.org/}.
Our experiments with these tools showed inconsistent measurements, especially for applications with a relatively small memory footprint of a few hundred megabytes.
Therefore, we measured the memory volume instead of the peak amount of memory.
Since the goal is to compare \petsc and \hyteg, this method is well suited.

\begin{figure}
  \includegraphics[width=\columnwidth]{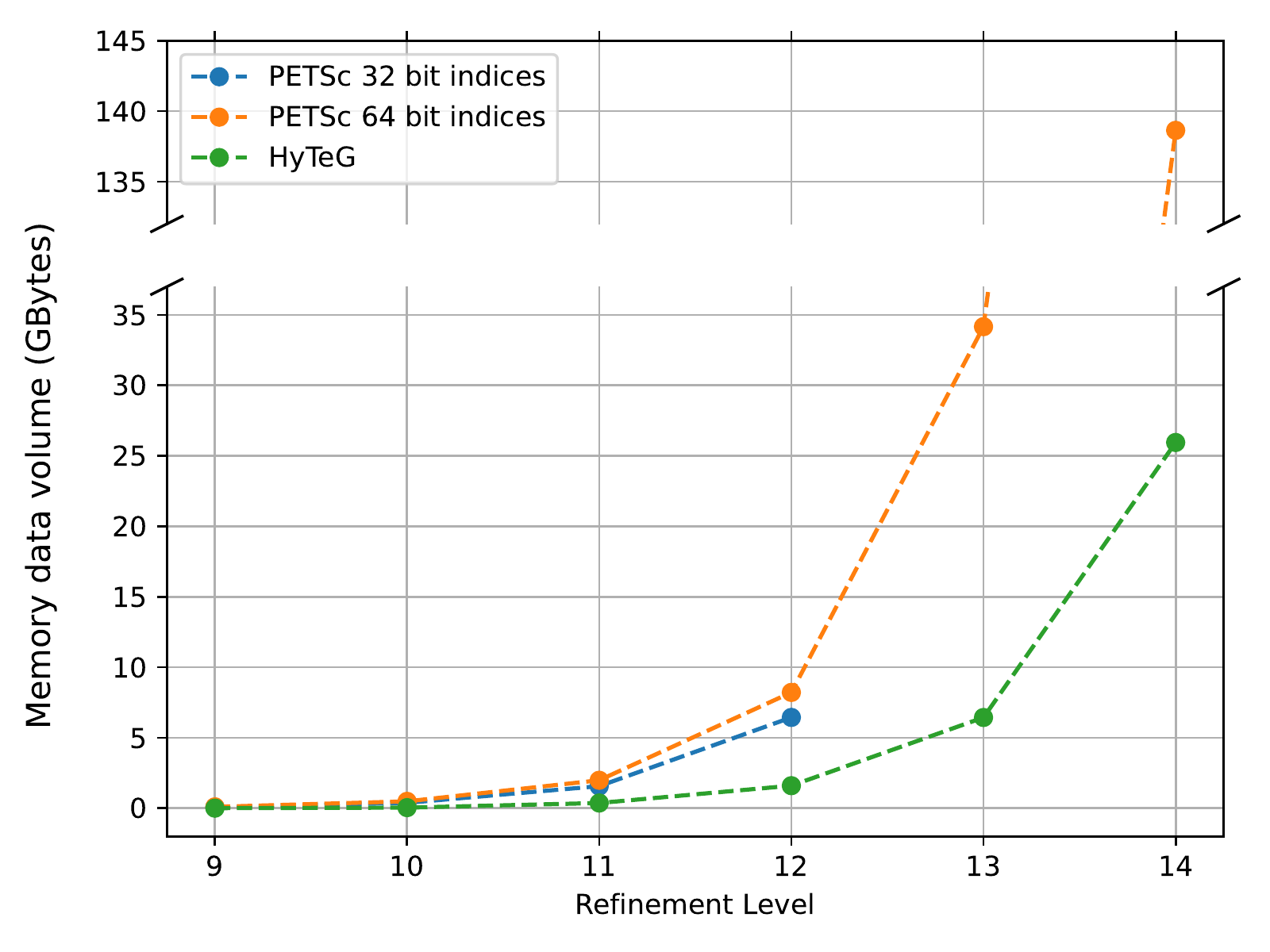}
  \caption{Memory data volume for the matrix-vector multiplication of \hyteg and \petsc for different refinement levels. Note that \petsc with 32-bit indices can only run up to refinement level 12. The difference converges to a factor of $\approx 5$ as expected.}
    \label{fig:petsc-vs-hyteg-mem}
\end{figure}

In the benchmark application, \hyteg first sets up the stencils and then transforms them into a PETSC matrix stored in a compressed format.
In the next step, the application performs a matrix-vector multiplication using the \petsc sparse-matrix and the \hyteg stencils while measuring the memory volume for both operations separately.
Figure~\ref{fig:petsc-vs-hyteg-mem} shows the difference between the two implementations.
One interesting fact is that if \petsc is built with 32-bit integers, it can only be used up to refinement level 12 since, at refinement level 13, there are more entries in the matrix than $2^{32}$.
Using 64-bit integers for indices further increases the memory footprint of \petsc.
Finally, above refinement level 14, the total memory limit of the system is reached with \petsc while \hyteg can utilize two steps more.
This shows that \hyteg is capable of performing a matrix-vector multiplication with up to  $8.6 \cdot 10^9$ \gls{dof} on a system with \SI{96}{\giga\byte}.

In summary, the benchmark application confirms the theoretical advantages of matrix-free methods concerning memory usage.
Since, in our case, the matrix-vector multiplication is memory bound, the % application's
performance is cross-checked by measuring the memory bandwidth and comparing it to the maximal memory bandwidth achievable.
The analysis with the \likwid tools revealed that the abstract \CPP implementation on refinement level 12 only achieves $\approx$ \SI{12.5}{\giga\byte\per\second}.
This bandwidth is far from the theoretical maximum, as shown in Table~\ref{tab:bandwidth}.
The following section explores how this gap can be closed using code generation to optimize performance.

%% file: codegen.tex
%!TEX root = performance.tex
%
\section{Code Generation}\label{sec:codegen}
This section will introduce how code generation is used to generate highly performant compute kernels from abstract \CC code.
The following shows a pseudo version of the abstract \CC kernel for the matrix-vector multiplication kernel that couples the \gls{dof} located at the vertices:
\begin{lstlisting}[language=C,basicstyle=\ttfamily\tiny]
void kernel-vertex-to-vertex( double * triangle_destination , double const * const triangle_soucrce , double const * const stencil_vtv, int level ){
for ( const auto& it : Iterator( level, 1 ) ){
  tmp = 0.0;
  for ( const auto direction : neighborsWithCenter ){
    tmp += stencil_vtv[stencilIndexFromVertex( direction )] *
           triangle_soucrce[indexFromVertex( level, it.x(), it.y(), direction )];
  }
  triangle_destination[indexFromVertex( level, it.x(), it.y(), stencilDirection::VERTEX_C )] = tmp;
}}
\end{lstlisting}
Code generation can be used in various ways to aid software development.
The idea is generally to specify the problem in an abstract and often simpler and more compact form.
Code generation allows for faster development and reduces the risk of errors
since less code must be developed and maintained.
Another advantage is that with a general problem specification, the generation framework can produce code adjusted to the targeted hardware architecture and take advantage of its specific features.

In \hyteg, we utilize code generation only for the time-consuming parts of the code in contrast to other approaches in which the whole framework, including the data structures, is generated.
The Exastenicls framework \cite{lengauer2014exa} employs whole program generation using a domain-specific language to generate advanced multigrid solvers.
Other examples are the FEniCSx \cite{ScroggsEtal2022} and the Firedrake \cite{Bercea2016Firedrake} projects, which generate code for the solution of partial differential equations using the finite element method.

For the generation of the compute kernels, we use the \pystencils \cite{bauer2019codegen} code-generation framework, which utilizes an embedded DSL in Python.
The general idea is to specify a stencil-based interpretation of the compute kernels and automatically transfer these to highly optimized \CC code.
\pystencils uses and extends the computer algebra package \sympy such that stencil kernels can be formulated symbolically.
The framework can then transform the symbolic representation into an abstract syntax tree (AST),
a tree representation of the code structure. % in an abstract formulation.
Based on the AST, \pystencils{} performs various optimizations,
such as eliminating common subexpressions.
Finally, \pystencils renders the AST into compilable \CC source code.
It is worth mentioning that other backends are also available to generate source code for different hardware like GPUs.
Once the desired \CC functions are generated, these can replace the original abstract \CC functions in the \hyteg framework.

The following shows an example of the \pystencils code to generate the kernel, which handles the matrix-vector multiplication for the \gls{dof} located at the vertices.
Additionally, the resulting \CPP code is shown to demonstrate the usefulness of \pystencils.
In contrast to the Python code, which is relatively easy to read, the \CPP kernel is highly error-prone, especially concerning indexing.

\begin{lstlisting}[language=Python]
triangle_source = VertexTriangleField('triangle_soucrce', const=True)triangle_destination = VertexTriangleField('triangle_destination')
vertex_to_vertex_stencil = StencilField('vtv', vertex_stencil_layout)
level = 10
neighbors = list()
for direction in vertex_stencil_layout:
    neighbors.append(vertex_to_vertex_stencil[direction] * triangle_source(direction))

update = sympy.Eq(triangle_destination((0, 0)), sum(neighbors))

kernel = create_kernel('kernel', [update], level)
print(generate_c_code(kernel))

\end{lstlisting}

\begin{lstlisting}[language=C,basicstyle=\ttfamily\tiny]
void kernel(double * _data_triangle_destination, double const * const _data_triangle_soucrce, double const * const _data_vtv)
{
const double xi_0= _data_vtv[2]; const double xi_1= _data_vtv[5]; const double xi_2 = _data_vtv[0]; const double xi_3= _data_vtv[3];
const double xi_4= _data_vtv[6]; const double xi_5= _data_vtv[1]; const double xi_6= _data_vtv[4];
for (int ctr_2 = 1; ctr_2 < 1024; ctr_2 += 1) {
for (int ctr_1 = 1; ctr_1 < 1024 - ctr_2; ctr_1 += 1) {
 const double xi_10 = xi_0*_data_triangle_soucrce[ctr_1 + 1026*ctr_2 - ((ctr_2*(ctr_2 + 1)) / (2)) - 1];
 const double xi_11 = xi_1*_data_triangle_soucrce[ctr_1 + 1026*ctr_2 - (((ctr_2 + 1)*(ctr_2 + 2)) / (2)) + 1025];
 const double xi_12 = xi_2*_data_triangle_soucrce[ctr_1 + 1026*ctr_2 - ((ctr_2*(ctr_2 - 1)) / (2)) - 1026];
 const double xi_13 = xi_3*_data_triangle_soucrce[ctr_1 + 1026*ctr_2 - ((ctr_2*(ctr_2 + 1)) / (2))];
 const double xi_14 = xi_4*_data_triangle_soucrce[ctr_1 + 1026*ctr_2 - (((ctr_2 + 1)*(ctr_2 + 2)) / (2)) + 1026];
 const double xi_15 = xi_5*_data_triangle_soucrce[ctr_1 + 1026*ctr_2 - ((ctr_2*(ctr_2 - 1)) / (2)) - 1025];
 const double xi_16 = xi_6*_data_triangle_soucrce[ctr_1 + 1026*ctr_2 - ((ctr_2*(ctr_2 + 1)) / (2)) + 1];
 _data_triangle_destination[ctr_1 + 1026*ctr_2 - ((ctr_2*(ctr_2 + 1)) / (2))] =
   xi_10 + xi_11 + xi_12 + xi_13 + xi_14 + xi_15 + xi_16;
}}}
\end{lstlisting}

%% file: 2D-ECM.tex
%!TEX root = performance.tex
%
\section{Performance Modeling}\label{sec:performance-model}

\subsection{Experiment Description}

\begin{figure}
  \centering
  \begin{subfigure}[t]{0.45\columnwidth}
      \includegraphics[width=\textwidth]{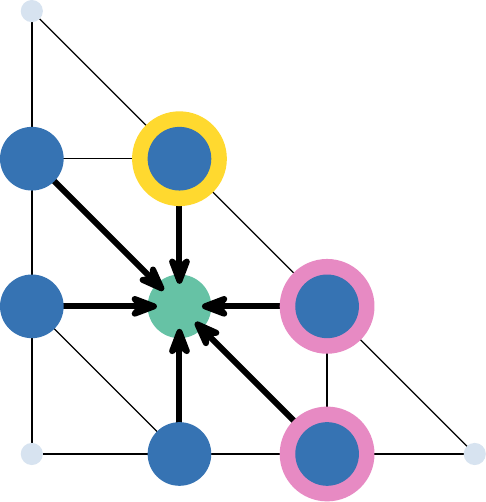}
      \caption{Vertex-to-Vertex Stencil}
      \label{fig:vtv}
  \end{subfigure}
  \hfill
  \begin{subfigure}[t]{0.45\columnwidth}
      \includegraphics[width=\textwidth]{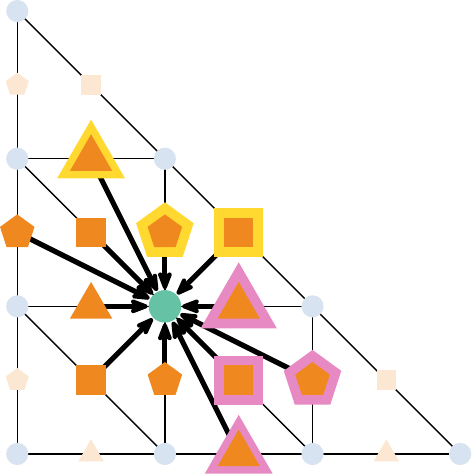}
      \caption{Edge-to-Vertex Stencil}
      \label{fig:etv}
  \end{subfigure}

  \begin{subfigure}[t]{0.45\columnwidth}
    \includegraphics[width=\textwidth]{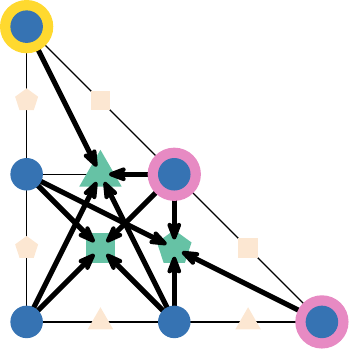}
    \caption{Vertex-to-Edge Stencil}
    \label{fig:vte}
  \end{subfigure}
  \hfill
  \begin{subfigure}[t]{0.45\columnwidth}
    \includegraphics[width=\textwidth]{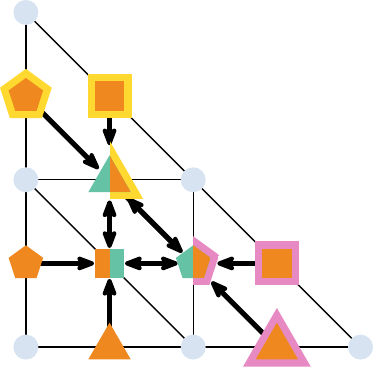}
    \caption{Edge-to-Edge Stencil}
    \label{fig:ete}
  \end{subfigure}
  \caption{The four stencils involved in the matrix-vector multiplication. The borders describe the availability in the cache.
  \\No borders: the entry was used in the last iteration and is still in the L1 cache.
  \\Yellow: the entry was never loaded before.
  \\Pink: the entry was used in a former iteration and might still be in the cache.}
  \label{fig:four-stencils}
\end{figure}

The profiling in Section~\ref{sec:app} revealed that the matrix-vector multiplication is the one part of the code that consumes most of the time, which is often the case in numerical simulation codes.
In this section, we only focus on the matrix-vector multiplication analyses.
However, \hyteg also applies code generation to other operations like vector addition or smoothers in the context of multigrid.
\hyteg uses functions and operators as basic build blocks.
A matrix-vector multiplication is equivalent to applying an operator to a function.
Therefore, the matrix-vector multiplication is also referred to as \verb-apply- in this publication.
In our implementation, the matrix-vector multiplication consists of four different kernels, as explained in Section~\ref{sec:comparing}.
Each kernel deals with a different combination of vertex and edge \gls{dof}.
Therefore, the resulting kernels are
\begin{itemize}
  \item Vertex-to-Vertex(Figure \ref{fig:vtv}),
  \item Edge-to-Vertex (Figure \ref{fig:etv}),
  \item Vertex-to-Edge (Figure \ref{fig:vte}),
  \item Edge-to-Edge (Figure \ref{fig:ete}).
\end{itemize}
Each of these kernels is analyzed separately in the corresponding section.
When iterating, we start at the lower left (the 90-degree angle of the triangle), and the horizontal direction (X) is the inner loop.
These kernels do not update the \gls{dof} located on the boundaries.
The boundary points are handled by dedicated interface primitives as explained in \anon{\cite{kohl2019hyteg}}.
We chose a minimum refinement level of seven to guarantee at least 100 entries in the longest row.
Otherwise, the loop overhead is observable, and the performance is limited.

\subsection{Vertex to Vertex Kernel}

The kernel that couples vertex and vertex \gls{dof} is analyzed first and is shown in Listing~\ref{lst:vtv}.
It is similar to a 9-point stencils kernel which contains all direct neighbors in two dimensions, including the diagonal ones.
The only difference is that the top right and bottom left entries are missing due to the triangular mesh.
Figure~\ref{fig:vtv} shows the stencil pattern.

\begin{lstlisting}[caption={Vertex-to-Vertex kernel},label=lst:vtv,frame=single,float,style=mystyle]
dstVertex[y][x] =
  srcVertex[y-1][x+1] * c0 + srcVertex[y  ][x+1] * c1 +
  srcVertex[y-1][x  ] * c2 + srcVertex[y  ][x  ] * c3 +
  srcVertex[y+1][x  ] * c4 + srcVertex[y  ][x-1] * c5 +
  srcVertex[y+1][x-1] * c6
\end{lstlisting}

Table \ref{tab:supermuc} describes the Intel Skylake architecture for which the kernel is analyzed.
As described in Section~\ref{sec:ecm-model}, the basic units in the ECM model are \acrfullpl{cl}.
We must determine the work unit, meaning how many operations are executed for data in a single \gls{cl}.
The Skylake CPU has a \gls{cl} size of \SI{64}{\byte} on all cache levels.
If double precision (\SI{8}{\byte}) is assumed, one work unit performs eight kernel iterations.

In the scalar case, the kernel executes seven multiplications and six additions for each iteration.
For a work unit of eight iterations, this leads to $(7+6)*8=104$ operations.
One AVX instruction operates on a vector length of \SI{32}{\byte}, equivalent to four doubles.
Therefore, 14 \verb-FMA AVX- instructions are needed to process one work unit iteration and $T_{OL} = 7 cy$.
However, these are only theoretical numbers that do not consider the CPU's hardware features, like the port utilization.
To obtain more realistic numbers, we use the IACA\cite{iaca} tool, which reports that it takes \SI{10}{\cycle} to execute one work unit.
The CPU can sustain two \verb-AVX load- and one AVX \verb-store- per cycle,
and the kernel needs 14 \verb-AVX LOADs- and two \verb-AVX STOREs-, which leads to $T_{nOL}=7 cy$.
Again, using IACA shows that practically \SI{8}{\cycle} are needed.

Next, the data transfer is analyzed.
For the target array \verb-dstVertex-, two CL transfers are needed throughout the whole memory hierarchy since every store miss leads to a write-allocate.
We assume that the array entries $a[x-1][y+1], a[x-1][y], a[x][y]$ and $a[x][y-1]$ are in the $L1$ cache since they have been used in the previous iteration.
These three entries are shown in Figure~\ref{fig:vtv} without a border and contribute only to $T_{nOL}$.
The entry $a[x][y+1]$ (yellow border in Figure~\ref{fig:vtv}) must always be loaded from the lowest memory level.
Up to this point, there are two CL loads and one CL store for each work unit.
Since the connection between the L1 and L2 cache has a half-duplex bandwidth of \SI{64}{\byte\per\second}, it takes three cycles to transfer three cache lines ($T_{L1L2} = 3$ \si{\cycle}).
For the transfer between L2 and L3, only the two cache line loads are considered since the store happens simultaneously due to full-duplex.
With the given bandwidth of \SI{16}{\byte\per\cycle} this results in 8 cycles ($T_{L2L3} = 8$ \si{\cycle}).

The analysis for the remaining entries $a[x+1][y]$ and $a[x+1][y-1]$ is more complex since it depends on the size of the leading dimension and whether these entries are still in the cache.
The \gls{LC}, as described in Section~\ref{sec:ecm-model}, can determine the cache behavior.
In this case, if three successive rows of the array can reside in a particular cache level, the \gls{LC} for that cache level is fulfilled.
Otherwise, the lower levels in the memory hierarchy have to be accessed.
For a certain cache level $k$, a number of entries $n$ and the cache Size $C_k$, this condition is $(3+1) \cdot n \cdot 8B < C_k$.
We add one line to account for the target array \verb-dstVertex-.
To simplify the calculations, we ignore that the row size decreases by one with each outer iteration.

The hierarchy of the \hyteg grids allows only for a finite set of array lengths depending on the level of refinement.
Therefore we can determine for which cache level the \gls{LC} is fulfilled for a particular refinement level.

\begin{table}
  \small
\begin{tabular}{ccccc}
  \toprule
  LC & levels & ECM model & prediction & prediction \\
  &  & cycles & cycles & \si{\giga\flop\per\second} \\
  \midrule
  L1 & 7-10  & $\{10 || 8 |3 | 8 | - \}$ & $\{10 \rceil 11 \rceil 19 \rceil - \rceil \}$ & $14.8$\\
  L2 & 11-14 & $\{10 || 8 |5 | 8 | 5.0 \}$ & $\{10 \rceil 13 \rceil 21 \rceil 26 \rceil \}$ & $10.8$\\
  \bottomrule

\end{tabular}
\caption{ECM model for the Vertex-to-Vertex kernel. ECM model and prediction are stated in CPU cycles (cy).}
\label{tab:ecmVV}
\end{table}

Table~\ref{tab:ecmVV} shows the ECM model for different refinement levels.
If the level is below eleven, the \gls{LC} for the L1 cache is fulfilled, and additionally, the whole data fits into the L3 cache.
The last column states the performance prediction for the highest level in the given range.

To calculate the predicted performance in \si{\giga\flop}, we multiply the number of flops by the CPU frequency and divide that through the predicted cycles.
In the case that the L1 \gls{LC} is fulfilled the prediction is $(8 \cdot \SI{13}{\flop} \cdot \SI{2.7}{\giga\hertz})/\SI{19}{\cycle}=\SI{14.78}{\giga\flop\per\second}$.

\begin{figure}
  \includegraphics[width=\columnwidth]{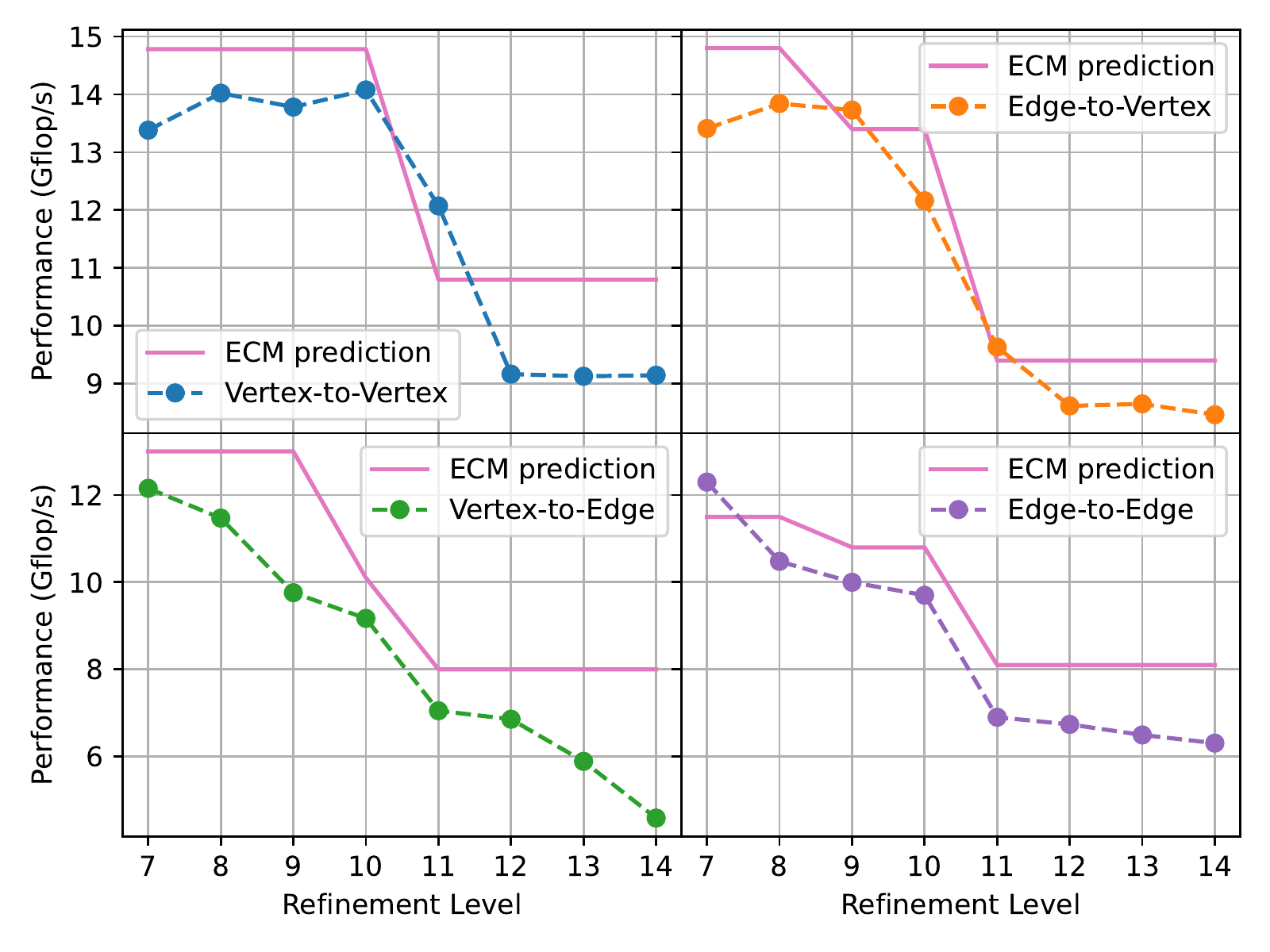}
  \caption{ECM model predictions and measurements for the different kernels analyzed in Section \ref{sec:performance-model}.}
    \label{fig:bench}
\end{figure}

For the calculations in Table~\ref{tab:ecmVV}, we assume that the \gls{LC} is constant for the whole kernel, which is not the case since we deal with triangular domains.
Therefore the actual measured performance is different, as shown in Figure~\ref{fig:bench}.
%Due to the small array sizes, the performance is lower for small numbers of refinement, which displays in the performance increase up to level 10.

%One can also see evidence for this explanation in the analysis of the performance counters, which are special counters inside the CPU capable of counting certain operations like FLOPs or cache misses.
%By utilizing these counters, it becomes apparent that the percentage of the vectorized floating-point operations compared to the scalar floating-point operations increases up to level 10.

Refinement level 11 shows a higher performance than the prediction of \SI{10.8}{\giga \flop \per \second}.
We explain this by constantly decreasing row sizes due to the triangular structure.
After a certain number of iterations, the rows will start to fit into the L1 cache again.
It is assumed that the \gls{LC} for the L1 cache is fulfilled for the upper half of the rows, which explains the performance overshoot compared to the prediction.
For level 12, the ratio of \emph{L1 \gls{LC}} reduces to bellow \SI{10}{\percent}, and therefore, the effect can be neglected.

%Meaning that 25\% of the code can be executed with \SI{12.04}{\giga \flop \per \second} and 75\% with \SI{10.8}{\giga \flop \per \second}.
%This leads to a performance prediction of $0.25 * \SI{14.78}{\giga \flop \per \second} + 0.75 * \SI{10.8}{\giga \flop \per \second} = \SI{11.8}{\giga \flop \per \second}$.

The prediction of \SI{10.8}{\giga \flop \per \second} is \SI{17}{\percent} above the actual measured performance of \SI{9}{\giga \flop \per \second}.
The results from how the ECM model was used in this publication do not consider the changes to the L3 cache with the introduction of the Skylake microarchitecture.
Details on these changes, introduced with Skylake, can be found in \cite{alappat2020intelcpu}.

\subsection{Edge-to-Vertex}

This section analyzes the kernel that couples the edge \gls{dof} to the vertex \gls{dof}.
Listing~\ref{lst:etv} and Figure~\ref{fig:etv} show the details.
Since the horizontal, vertical, and diagonal edge DoFs are split into separate arrays, there are three loads and one store stream.

\begin{lstlisting}[caption={Edge-to-Vertex kernel},label=lst:etv,frame=single,float]
vertex[y][ x] =
  edgeX[y+1][x  ] * c0   + edgeX[y  ][x  ] * c1 +
  edgeX[y  ][x-1] * c2   + edgeX[y-1][x  ] * c3 +

  edgeY[y  ][x-1] * c4   + edgeY[y  ][x  ] * c5 +
  edgeY[y-1][x  ] * c6   + edgeY[y-1][x+1] * c7 +

  edgeXY[y  ][x-1] * c8  + edgeXY[y  ][x  ] * c9 +
  edgeXY[y-1][x-1] * c10 + edgeXY[y-1][x  ] * c11;
\end{lstlisting}

Each iteration of the kernel executes twelve MULT and eleven ADD instructions.
Normalizing these numbers to an entire cache line yields 24 MULT and 22 ADD.
With FMA AVX, the theoretical throughput is at 12 cycles, but using the IACA tool, the prediction is $24$ cycles per iteration, leading to $T_{OL} = 24 cy$.

Concerning the data transfer from the L1 cache, there are twelve load
streams and one store stream resulting in 2 AVX STOREs and 24 AVX
LOADs per cache line. Therefore $T_{nOL} = 12$, which IACA also confirmed.

For the data transfer between the cache level, the writing to the target array \verb-vertex- causes the store and load of one cache line due to write-allocate.
The \verb-edgeX-, \verb-edgeY-, and \verb-edgeXY- arrays depend again on the \gls{LC}.
In the best case, only one entry per array (yellow border in Figure~\ref{fig:etv}) needs to be fetched from memory summing up to a total of three.
The five entries without a border in Figure~\ref{fig:etv} are always available in the L1 cache.
The other four entries reside in the L1 cache as long as the corresponding \gls{LC} is fulfilled.
Figure~\ref{fig:etv} shows these entries with a pink border.
In this case, the predictions from the ECM model are $T_{L1L2} = 5 cy$ and $T_{L2L3} = 16 cy$.
This condition holds if the refinement level is below ten, and this also means that the entire data resides in the L3 cache, meaning there is no traffic between the L3 and main memory.

If the \gls{LC} is violated, four additional loads exist between each cache level.
Interestingly, due to the stencil's shape for the \verb-edgeY- and \verb-edgeXY- arrays, there is only one additional entry, but for the \verb-edgeX- array, there are two.
This is because the kernel touches both \verb-y+1- and \verb+y-1+, whereas in the other cases, only \verb-y+1- is needed.

From level eleven onwards, there is also traffic between the L3 cache and the main memory.
Since there are three loads and one store stream, we use a transfer time of \SI{2.0}{\cycle} per cache line from Table~\ref{tab:bandwidth}.

\begin{table}
  \small
\begin{tabular}{ccccc}
  \toprule
  LC & levels & ECM model & prediction & prediction \\
  &  & cycles & cycles & \si{\giga\flop\per\second} \\
  \midrule
  L1 & 7-8   & $\{24 || 12 |5 | 16 | - \}$ & $\{24 \rceil 24 \rceil 33 \rceil - \rceil \}$ & $15.1$\\
  L2 & 9-10 &  $\{24 || 12 |9 | 16 | 8\}$ & $\{24 \rceil 24 \rceil 45 \rceil 53 \rceil \}$ & $13.4$\\
  L2 & 11-14 & $\{24 || 12 |9 | 24 | 8\}$ & $\{24 \rceil 24 \rceil 45 \rceil 53 \rceil \}$ & $9.4$\\
  \bottomrule

\end{tabular}
\caption{ECM model for Edge-to-Vertex kernel.}
\label{tab:ecm-etv}
\end{table}

Figure~\ref{fig:etv} show the measurements for this kernel, which are in good accordance with the predictions.

\subsection{Vertex-to-Edge Kernel}

Now the kernel coupling from vertex DoFs to edge DoFs is analyzed, shown in Listing~\ref{lst:vte} and Figure~\ref{fig:vte}.
In contrast to the Vertex-to-Vertex kernel, it writes to three data targets (\verb-edgeX,edgeY,edgeXY-).

\begin{lstlisting}[caption={Vertex-To-Edge kernel},label=lst:vte,frame=single,float]
edgeX[y][ x] =
  vertex[y+1][x-1] * c0  + vertex[y  ][x  ] * c1 +
  vertex[y  ][x+1] * c2  + vertex[y-1][x+1] * c3;

edgeXY[y][ x] =
  vertex[y+1][x  ] * c4  + vertex[y+1][x+1] * c5 +
  vertex[y  ][x  ] * c6  + vertex[y  ][x+1] * c7;

edgeY[y][ x] =
  vertex[y+1][x-1] * c8  + vertex[y+1][x  ] * c9 +
  vertex[y  ][x  ] * c10 + vertex[y  ][x+1] * c11;
\end{lstlisting}

This kernel performs four multiplications and three additions for all three target arrays in each iteration.
A total of twelve MULT and nine ADD instructions.
Using vectorization and fused multiply-add, this reduces to 12 FMA AVX instructions for four iterations and 24 FMA AVX instructions for one work unit.
Theoretically, two FMA AVX instructions can be executed per cycle, suggesting that processing an entire cache line needs 12 cycles.
However, an analysis of the kernel using the IACA tool shows that the throughput is slightly higher $T_{OL}= 14.8cy$.

The kernel writes into three different array locations and reads from twelve.
However, \verb+[x][y]+, \verb-[x+1][y]- and \verb-[x][y+1]- are used three times, and therefore only six actual loads are needed.
IACA predicts that these loads and stores take 12 cycles ($T_{nOL}=12cy$).

Now the data transfers between the cache levels are analyzed.
There are three store streams and three load streams due to write-allocate for the target arrays \verb-edgeX-, \verb-edgeY-,\verb-edgeXY-.
At least one CL transfer is needed for the source array \verb-srcVertex-.
Figure~\ref{fig:vte} show this point with a yellow border.
Similar to the Vertex-to-Vertex kernel, there are two entries in the source array (pink border) where the \gls{LC} determines from which level of the memory hierarchy they are fetched.
If the \gls{LC} for the L1 cache is fulfilled $T_{L1L2} = 7 cy$ and $T_{L2L3} = 16$.
Once the L1 \gls{LC} is violated $T_{L1L2}$ increases to $9 cy$, and in the case that the L2 \gls{LC} is also violated $T_{L2L3}$ becomes $24 cy$.
Remember that only the load streams are considered for $T_{L2L3}$ since this connection is full-duplex and can read and write simultaneously.

For the transfer between the L3 cache and main memory, we use the bandwidth from Table \ref{tab:bandwidth}.
Since there is one load and three store streams to memory, $T_{L3MEM}$ is $\SI{11.6}{\cycle}$.

Table~\ref{tab:ecmVE} displays the ECM model predictions for the kernel.
The ECM predictions here are in good accordance except for the higher levels, where the prediction and the measurements diverge.
We attribute this to the fact that at refinement level 15, the \gls{LC} for the L3 cache would be violated, which already shows at smaller sizes.
Additionally, the ratio of three stores to one load stream reduces the performance.
Table~\ref{tab:bandwidth} also displays this characteristic, where the respective benchmark reaches the lowest bandwidth.

\begin{table}
  \footnotesize
\begin{tabular}{ccccc}
  \toprule
  LC & levels & ECM model & prediction & pred. \\
  &  & cycles & cycles & \si{\giga\flop\per\second} \\
  \midrule
  L1 & 7-9   & $\{14.8 || 12 |7 | 16 | - \}$ & $\{14.8 \rceil 19 \rceil 35 \rceil - \rceil \}$    & $13.0$\\
  L2 & 10    & $\{14.8 || 12 |9 | 24 | - \}$ & $\{14.8 \rceil 21 \rceil 45 \rceil - \rceil \}$   & $10.1$\\
  L2 & 11-14 & $\{14.8 || 12 |9 | 24 | 11.6 \}$ & $\{14.8 \rceil 21 \rceil 45 \rceil 56.6 \rceil \}$ & $8.0$\\
  \bottomrule
\end{tabular}
\caption{ECM model for Vertex-to-Edge kernel.}
\label{tab:ecmVE}
\end{table}
\subsection{Edge-to-Edge}

The last kernel to be analyzed is the one coupling edge and edge \gls{dof} shown in Listing~\ref{lst:ete} and Figure~\ref{fig:ete}.
This kernel uses three load streams and three store streams.

\begin{lstlisting}[caption={Edge-to-Edge kernel},label=lst:ete,frame=single,float]
dstEdgeX[y][ x] =
  srcEdgeX [y  ][x  ] * c0 + srcEdgeY [y  ][x  ] * c1 +
  srcEdgeY [y-1][x+1] * c2 + srcEdgeXY[y  ][x  ] * c3 +
  srcEdgeXY[y-1][x  ] * c4;
dstEdgeY[y][x] =
  srcEdgeY [y  ][x  ] * c5 + srcEdgeX [y  ][x  ] * c6 +
  srcEdgeX [y+1][x-1] * c7 + srcEdgeXY[y  ][x  ] * c8 +
  srcEdgeXY[y  ][x-1] * c9;
dstEdgeXY[y][x] =
  srcEdgeXY[y  ][x  ] * c10 + srcEdgeX [y  ][x  ] * c11 +
  srcEdgeX [y+1][x  ] * c12 + srcEdgeY [y  ][x  ] * c13 +
  srcEdgeY [y  ][x+1] * c14;
\end{lstlisting}

For each group of edge DoFs, five MULT and four ADD instructions per iteration need to be performed.
Like before, there are eight iterations when normalizing to an entire cache line, leading to $5 \cdot 3 \cdot 8 = 120$ multiplications and $4 \cdot 3 \cdot 8=96$ additions.
Within one FMA AVX instruction, a maximum of four MULT and four ADD instructions are executed, reducing the total operations to 30 FMA AVX instructions.
The IACA tool reports that it takes 20 cycles to process one cache line for this kernel which is higher than the theoretical value of 15 cycles since the Skylake architecture is theoretically capable of performing 2 FMA
AVX instructions per cycle ($T_{OL} = 20 cy)$.

Looking at the data traffic between the L1 cache and the registers, three writes, and nine loads (three per group) are necessary.
Since one AVX load is equivalent to four double loads, this results in six AVX STOREs and 18 AVX LOADs per cache line.
IACA shows that it takes ten cycles, which leads to $T_{nOL} = 10$

For the data traffic within the cache hierarchy, the \verb-dstEdgeX-, \verb-dstEdgeY-, and \verb-dstEdgeXY- arrays cause one store and one load stream each.
For the \verb-srcEdgeX-,\verb-srcEdgeY-, and \verb-srcEdgeXY- arrays, each array has one entry that was never touched, causing one load stream throughout the whole cache hierarchy.
The last iteration used one entry of each array, meaning these entries still reside in the L1 cache.
For the remaining three entries, the \gls{LC} determines the location in the memory hierarchy.
Figure~\ref{fig:ete} shows the entries in the L1 cache without a border, the new entries with a yellow border, and the entries that are possibly in the cache with a pink border.
If the L1 layer condition is fulfilled the ECM reads $T_{L1L2} = 9 cy$ and $T_{L2L3} = 24 cy$.
The ECM model for the different refinement levels is presented in Table~\ref{tab:ecm-ete}.

Above level eight, the L1 \gls{LC} no longer holds and $T_{L1L2}$ increases to \SI{12}{\cycle}.
Until level ten, the entire data can also be kept in the L3 cache, meaning there is no data traffic to memory.
From level eleven on, the memory traffic needs to be considered.
We used the bandwidth results for the \likwid benchmark with one load and one store stream since it has the same ratio.
The transfer time for one cache line from main memory to L3 cache is, therefore, \SI{2.5}{\cycle{}}
The six streams that are required result in a prediction of $T_{L3MEM}$ being \SI{15}{\cycle}.

\begin{table}
\small
\begin{tabular}{ccccc}
  \toprule
  LC & levels & ECM model & prediction & prediction \\
  &  & cycles & cycles & \si{\giga\flop\per\second} \\
  \midrule
  L1 & 7-8     & $\{20 || 10 |9  | 24 | - \}$   & $\{20 \rceil 19 \rceil 43 \rceil - \rceil \}$ & $11.5$\\
  L2 & 9-10   & $\{20 || 10 |12 | 24 | 15 \}$ & $\{20 \rceil 22 \rceil 46 \rceil 61 \rceil \}$ & $10.8$\\
  L2 & 10-14 & $\{20 || 10 |12 | 24 | 15\}$ & $\{20 \rceil 22 \rceil 46 \rceil 61 \rceil \}$ & $8.1$\\
  \bottomrule
\end{tabular}
\caption{ECM model for the Edge-to-Edge kernel}
\label{tab:ecm-ete}
\end{table}

Figure~\ref{fig:ete} compares the measurements with the predictions, which match very well.
The fact that on refinement level seven, the whole dataset nearly fits into the L2 cache explains this point being above the prediction of the ECM model.

%% file: multiCore.tex
%!TEX root = performance.tex
\section{Multi-Core Scaling}\label{sec:multicore}

This section analyzes the scaling of all four kernels when executing on a single node.
When using the ECM model to predict the performance for a specific number of cores, one has to differentiate between exclusive and shared resources within the CPU.
All compute units, registers, the L1, and the L2 cache are exclusive for each core.
However, the L3 cache and the main memory are shared amongst different cores.
Therefore, we expect perfect scaling if the data fits in the L2 cache.

\begin{figure}
  \includegraphics[width=\columnwidth]{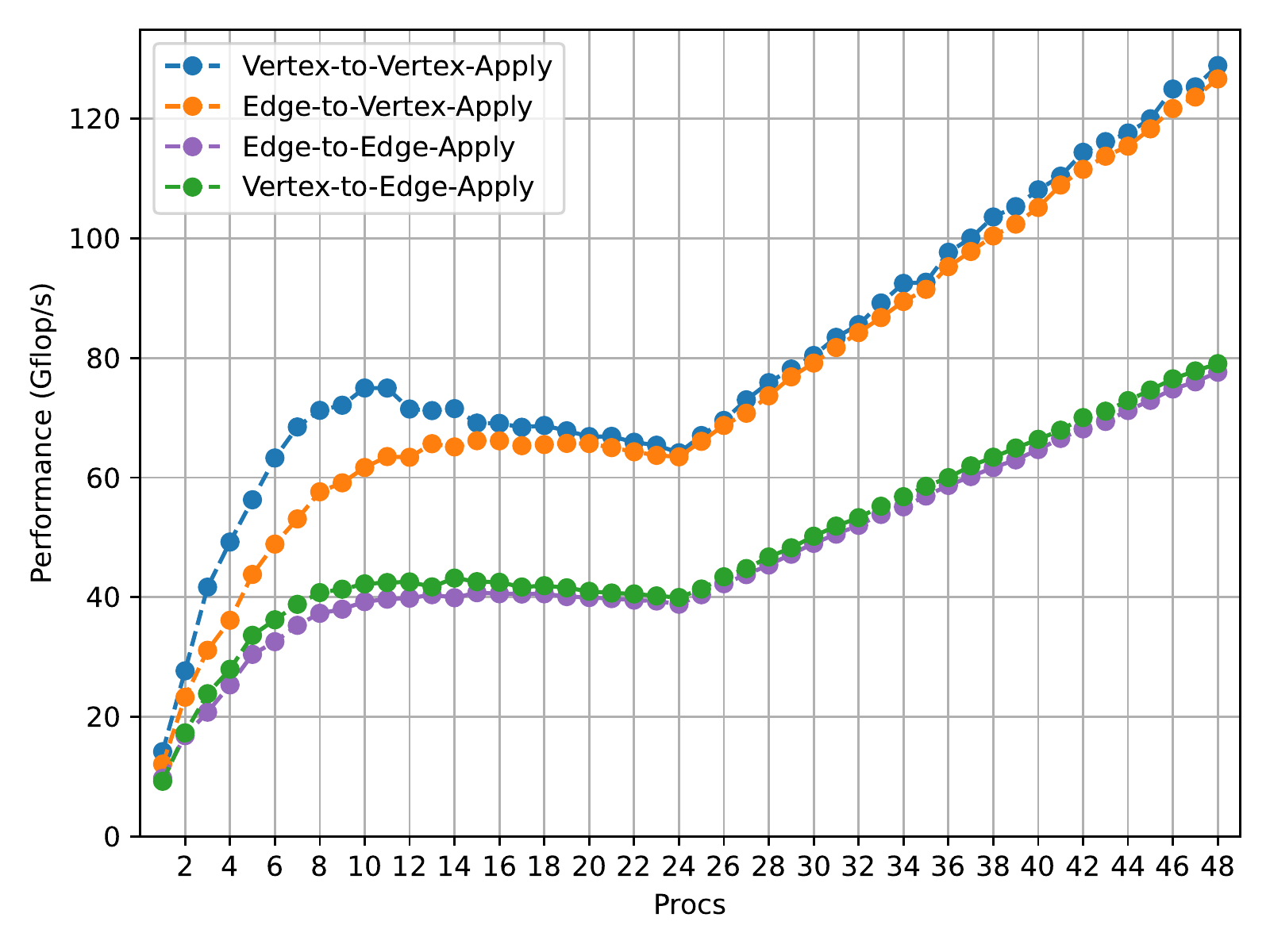}
  \caption{Intra-node scaling of the four different kernels to the full socket of one SuperMUC-NG node containing two sockets. The first 24 processes are scheduled onto one socket before the second socket is used. This scheduling explains the saturation at 24 processes due to the saturation in main memory.}
    \label{fig:scaling_level8}
\end{figure}

Figure~\ref{fig:scaling_level8} shows weak scaling of the four kernels from 1 to 48 processes for refinement level 8.
Weak scaling means the application adds a new triangle for each process to the mesh.
The scheduling is chosen to fill one socket before using the second socket.
As expected, the scaling up to six processes is linear since all the data can be kept in the L2 cache.
Beyond this, the shared L3 and main memory cause a saturation up to the entire socket containing 24 processes.

The \verb+Vertex-to-Vertex-Apply+ kernel shows a particular behavior.
After ten processes, the performance decreases because the shared L3 cache can no longer hold the whole mesh.
In other words, the size of the L3 cache available per core decreases when the number of processes increases.

Once processes utilize the second socket, the performance increases linearly.
The scheduling of the processes causes this characteristic.
Since the first socket is already fully saturated, the performance can only increase by $1/24$ because new processes have to wait for all 24 processes on the other socket. Overall this scaling behavior follows the characteristic of memory-bound kernels in general.

%git commit: 760a19c8; ICC 2019.1; SUPERMUC-NG

%% file: conclusion.tex
%!TEX root = performance.tex
%
\section{Conclusion and Outlook}\label{sec:outlook}
This paper analyses the differences between matrix-free and sparse matrix versions of matrix-vector multiplication.
One clear advantage of the matrix-free version is the reduced memory footprint.
In particular, the article compares the standard HPC library \petsc against the recently developed \hyteg framework concerning their memory requirements.
Here, \hyteg could reduce memory consumption significantly when the matrix-free methods were employed.
Furthermore, \petsc must be compiled with 64-bit integer indices to reach the same number of global \gls{dof} as \hyteg since the default 32-bit indices would overflow at refinement level 12.
Section~\ref{sec:codegen} shows how \hyteg uses code generation techniques to fully utilize the memory bandwidth.

The paper demonstrates how an elaborate exercise in performance modeling can be used to determine theoretical peak performance.
This analysis is based on the ECM model.
The article demonstrates that all analyzed kernels can reach the ECM-predicted performance limit when using code, the generation techniques of \hyteg.
A first single-core comparison of the complete matrix-vector multiplication of \hyteg with \petsc on the SuperMUC-NG system showed a speedup of 5.3 in favor of \hyteg.
However, more analyses would be required to evaluate this difference further.

The article thus presents a detailed performance analysis of the most critical kernels relevant for 2D simulations with the \hyteg framework.
However, as discussed in Section~\ref{sec:performance-model}, the four kernels needed to perform one complete matrix-vector multiplication are currently independent routines.
As such, they cause an overhead in memory transfer.
Merging these four kernels into a single one will accelerate the execution further.
The same code generation techniques as already used could significantly reduce the work.
However, the surrounding data structures need extensive adjustments to support the changed interface.
In the limit, this loop merge could improve the performance of the whole matrix-vector multiplication by another factor of four, provided that data can be reused entirely and the memory bandwidth is still the limiting performance factor.

Clearly, the next significant step is to extend the performance analysis to 3D \hyteg-meshes.
The transition from 2D to 3D leads to much more complicated stencil patterns, especially for \gls{dof} located at the edges.
A comprehensive study will be the topic of future research.

%% file: acknowledgments.tex
%!TEX root = performance.tex

\begin{acks}
The authors gratefully acknowledge the Gauss Centre for Supercomputing e.V.
(\url{https://www.gauss-centre.eu}) for funding this project by providing computing
time on the GCS Supercomputer SuperMUC-NG at Leibniz Supercomputing Centre (\url{https://www.lrz.de}).

Additional thanks to the HPC group at the Erlangen Regional Computing Centre (RRZE) for providing reliable test environments.

The development of \hyteg was made possible through the project TerraNeo \cite{bauer2020terraneo} funded by the German Priority Programme SPPEXA.

\end{acks}